# Programmable and scalable on-chip WDM processing platform enabled by cascaded FSR-free resonators

**Boshu Sun[1,2†], Haojie Zhu[1,2†], Ying Sun[3], Kunhao Lei[3], Zijia Wang[3], Kangjian Bao[1,2], Weiqi Lu[1,2], Yuhao Fang[1,2], Yuting Ye[1,2], Danyang Hao[1,2], Qingyan Deng[1,2], Zequn Chen[1,2], Yingchun Wu[1,2], Maoliang Wei[4], Wei Cao[4], Xu Sun[4], Qicheng Zhang[1,2], Hongtao Lin[3*], William Shieh[1,2,5*], Lan Li[1,2,5*]**

[1] *Zhejiang Key Laboratory of 3D Micro/Nano Fabrication and Characterization, Department of Electronic and Information Engineering, School of Engineering, Westlake University, Hangzhou 310030, China*

[2] *Institute of Advanced Technology, Westlake Institute for Advanced Study, Hangzhou 310024, China*

[3] *State Key Laboratory of Brain-Machine Intelligence, College of Information Science and Electronic Engineering, Zhejiang University, Hangzhou 310027, China*

[4] *Crealights Technology Co., Ltd.*

[5] *Westlake Institute for Optoelectronics, Fuyang, Hangzhou 311400, China*

† These authors contributed equally to this work.

hometown@zju.edu.cn, shiehw@westlake.edu.cn, lilan@westlake.edu.cn

## Abstract

Wavelength-division multiplexing (WDM) is pivotal for expanding the capacity and functionality of photonic systems, from communications to computing. However, on-chip implementation for broadband operation is fundamentally limited by the narrow free spectral range (FSR) and inefficient drop transmission of conventional optical microcavities. Here, we transcend this limit with a scalable integrated WDM processing platform based on cascaded dual-sided coupled Fabry-Pérot add-drop resonators. This unit achieves a record-large FSR-free bandwidth (>300 nm) and low insertion loss (< 0.5 dB), while enabling independent manipulation of monochromatic light in both the wavelength and intensity domains. Leveraging these performances, we architect programmable cores with up to 20 channels in a single bus waveguide to demonstrate application versatility by configuring as: a high-capacity on-chip WDM communication fabric (2.4-Tbps single-link WDM transmission and O-to-C-band multi-subband (de)triplexing), a reconfigurable optical processor (broadband and hitless wavelength-selective switching), and a parallel computing accelerator (theoretically capable of 1.28-TOPS convolution operations, 16-channel modulation). This work establishes a unified and versatile WDM processing platform that bridges high-speed optical communication with in-line computing, charting a scalable and backward-compatible path toward petabit-per-second-class links and hundreds of TOPS of on-chip computation, resolving a critical bottleneck for next-generation photonic systems.

## Introduction

The exponential growth of global data traffic, driven by artificial intelligence (AI) and cloud computing, is fundamentally re-engineering the hierarchy of information processing, from long-haul networks down to short-reach even chip-level interconnects for high-bandwidth-density optical communication[1] and ultra-low-latency computation[2]. Within this evolving landscape, wavelength-division multiplexing (WDM) serves as a pivotal architectural technology, leveraging its wavelength dimension to enable scalable, parallel photonic systems essential for high-density interconnects[3,4], next-generation passive optical networks (PONs)[5,6], and ultra-short-reach optical accelerators. This spectral parallelism supports a diverse array of critical functions, including high-capacity communication, reconfigurable optical switching, and analog optical processing, all achieved on a unified photonic circuit platform [7-9]. The transformative potential of WDM technology is further amplified by the advent of integrated optical frequency combs, particularly chip-scale microcombs[10-12]. These sources generate dozens to hundreds of coherent, equidistantly spaced frequency lines, creating powerful synergies with WDM that unlock unprecedented system capacity and spectral efficiency for petabit-per-second links[13-16] and massively parallel processing[17,18]. However, fully exploiting this “comb-plus-WDM” paradigm requires more than a light source; it demands a programmable photonic core capable of manipulating this vast, continuous spectral resource across wideband ranges[19].

Translating this vision into a practical, scalable, and programmable platform is hindered by a fundamental device-level bottleneck, as developing a filtering unit that concurrently excels in broadband operation, low insertion loss, high reconfigurability, and scalability remains a significant challenge. This presents fundamental trade-offs for conventional on-chip WDM building blocks. Silicon photonic microring resonators (MRRs), despite their compactness, low power consumption, and exceptional wavelength selectivity, are widely utilized in on-chip wavelength routers[20-22], optical modulators[23,24], and neuromorphic processors[25-27]. However, MRRs are subject to scalability limitations arising from the trade-off between their inherent free spectral range (FSR) and optical loss[28]. FSR, fundamentally limited by the resonator’s optical path length, is critically mismatched with the overall spectral coverage of multi-line sources like microcombs and constrains the wavelength span of WDM. While smaller-radius resonators can theoretically extend the FSR and improve spectral efficiency, these are practically limited by increased bending loss and severe crosstalk. Consequently, scaling MRR-based WDM systems beyond 16 channels per lane remains challenging[29], and achieving closer channel spacing for higher density often limits single-wavelength data rates to below 50 Gbps due to bandwidth requirements for crosstalk suppression[30]. These combined constraints render MRRs insufficient for extensive multi-wavelength parallelism and high data rates required for future optical systems. Alternative non-resonant multiplexers, such as arrayed waveguide gratings (AWGs)[31] and Mach-Zehnder interferometers (MZIs)[32], provide wide, flat-top passbands for multi-wavelength systems but are inherently static and bulky[31,33]. Programmable photonic integrated circuits based on MZIs can synthesize arbitrary linear functions, enabling reconfigurable filters and optical neural network accelerators[18,26,34,35]; however, their extensive footprint and static power consumption severely limit scalability. To directly attack the FSR bottleneck, various engineered resonators have been explored. Both the vernier effect-based MRRs of non-integer ratios and contra-directional couplers (CDCs)-assisted MRR[36] can achieve nearly FSR-free operation, but precise resonant wavelength alignment increases their control complexity and power consumption[37]. A notable calibration-free approach used the photonic bandgap of a single-side-coupled photonic crystal Fabry-Pérot (F-P) resonator to achieve FSR-free filtering[38]. However, its conversion to a practical add-drop configuration incurred a prohibitive 6 dB loss induced by a bidirectional coupler[39]. As another key function of a reconfigurable WDM system, intensity tuning via conventional dynamic modulation methods (e.g., free-carrier injection) further exacerbates optical loss[37]. Thus, while various paths relax specific constraints, each introduces critical trade-offs in loss, control overhead, or functional utility. A low-loss, FSR-free, calibration-free add-drop resonator that also enables dynamic spectral management remains a long-standing challenge, hindering the development of a truly scalable and broadband on-chip WDM processing platform.

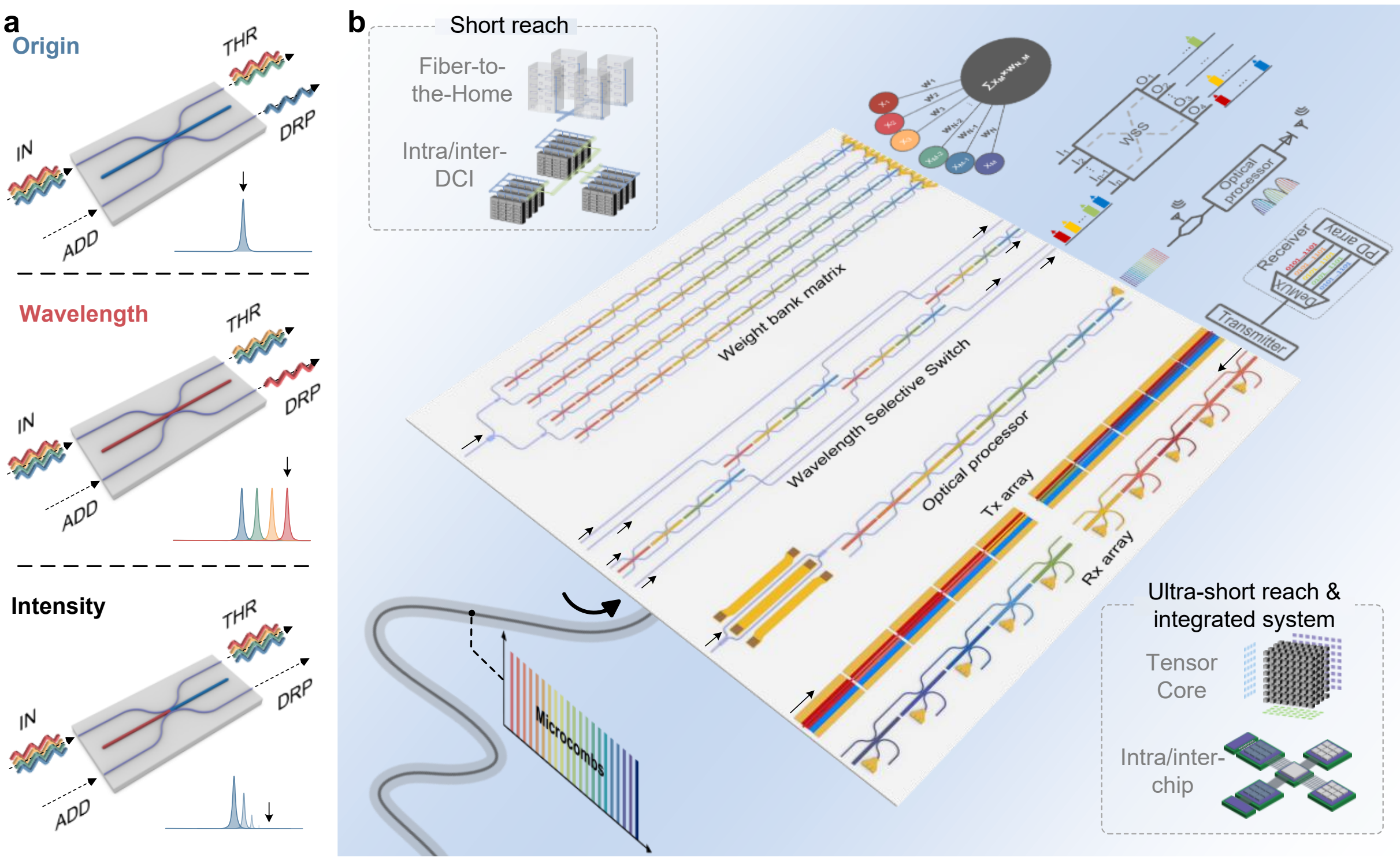

**Fig. 1 Proposed DC-FPR empowering various WDM applications.** **a** Conceptual diagram of the proposed DC-FPR unit, highlighting its flexible wavelength and intensity tuning capabilities. **b** Conceptual diagram for diverse on-chip WDM applications enabled by the DC-FPR platform, including data transmission, optical signal processing, wavelength routing/switching, and photonic acceleration. These systems operate across scales from short-reach to intra/inter-chip links and integrated optical tensor-core accelerators, leveraging microcomb sources on silicon photonic chips (DCI: data center interconnect; FttH: fiber-to-the-home).

Here, we bridge this gap by introducing a novel programmable on-chip WDM processing platform utilizing cascaded dual-sided coupled Fabry-Pérot resonators (DC-FPRs) as its add-drop filter building block. A conceptual illustration of this DC-FPR unit, highlighting its flexible wavelength- and intensity-tuning capabilities, is provided in Fig. 1a. Unlike MRRs or other conventional approaches, this DC-FPR architecture addresses the fundamental trade-off by simultaneously achieving an FSR-free bandwidth exceeding 300 nm and an insertion loss below 0.5 dB per unit. This breakthrough enables the deep cascading of dozens of units into a reconfigurable network that provides continuous access to the broadband spectrum. We demonstrate this platform as a universal "spectral engine" that unifies the hierarchy of optical systems by providing programmable control over the optical spectrum. This versatile DC-FPR platform, enabled by integrating microcomb sources on silicon photonic chips, as conceptually depicted in Fig. 1b, supports diverse on-chip WDM systems across various scales, from short-reach interconnects to intra- and inter-chip optical links and integrated optical tensor-core accelerators. Leveraging its inherent performance, we reconfigure the chip architecture to realize: (i) a high-capacity on-chip WDM communication fabric capable of 2.4-Tbps single-link data transmission and O-to-C-band multi-subband (de)triplexing; (ii) a reconfigurable and broadband optical processor functioning as a hitless wavelength-selective switch; and (iii) a parallel photonic accelerator delivering a theoretical throughput of 1.28 TOPS for matrix operations. These diverse implementations validate a fundamental transition from application-specific devices to a versatile on-chip WDM platform based on our designed cascaded filter units, thereby enabling scalable technologies for future optical interconnects and computing.

# Results

### Principle and characterization of DC-FPR

The proposed DC-FPR consists of two side-coupled bus waveguides and an F-P resonator formed by two Bragg grating-based mode conversion reflectors (MCRs), as shown in Fig. 2a. Unlike whispering-gallery-mode cavities such as MRR, a conventional F-P resonator operates as a standing-wave resonator, which inherently introduces 6-dB loss in the drop port (Supplement Note 1). To achieve high drop efficiency, we introduce $\pi$-phase-shifted MCRs into the F-P resonator, featuring a half-period shift between the grating teeth on the two sides of the waveguide. The central multimode waveguide, laterally coupled to single-mode bus waveguides, forms asymmetric mode directional couplers (AMDCs) that deliberately break modal symmetry. By engineering the waveguide widths and gaps to satisfy the phase-matching condition between the fundamental transverse-electric mode of the bus waveguide ($TE_{0_b}$) and the first-order mode of the multimode section ($TE_1$), the AMDC enables efficient, mode-order-selective power transfer. Consequently, the optical field evolves along the cascaded pathway: $TE_{0_b}$ (input) → $TE_1$ (multimode) → $TE_{0_r}$ (resonator) → $TE_1$ (multimode) → $TE_{0_b}$ (drop port), thereby completing the add-drop functionality without back-reflection into the input channel. Due to phase mismatch in the reverse direction, light leakage to other ports is suppressed, thereby significantly improving the drop efficiency. Furthermore, the grating parameters are engineered to tailor the mirror bandwidth such that the FSR-free condition ($FSR_{F\text{-}P}>BW_{MCR}/2$) is satisfied, effectively suppressing out-of-band resonant peaks and ensuring single-peak operation (Supplement Note 2). This calibration-free design eliminates the need for additional thermal or electrical tuning, reducing fabrication complexity and power consumption.

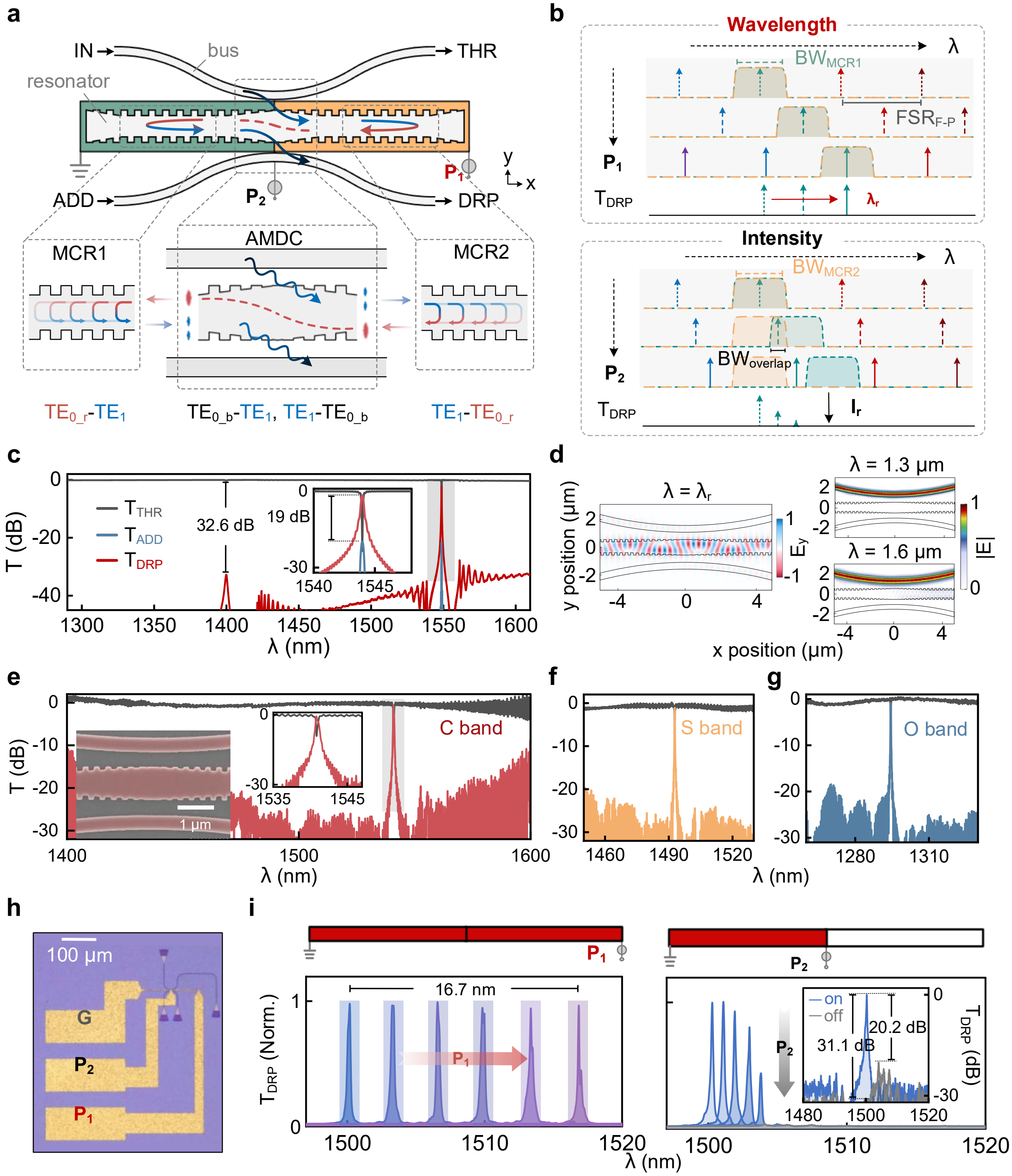

**Fig. 2 Principle and characterization of the versatile DC-FPR unit. a** Operation schematic of FSR-free add-drop bandpass filtering via mode coupling and conversion in the DC-FPR. High drop efficiency is achieved by combining MCRs that convert $TE_{0_r}$ to $TE_1$ mode with an AMDC. By utilizing this conversion for phase-matched directional coupling, reverse-phase mismatch suppresses light leakage to the IN and ADD ports. (green: spectral tuning heater of MCR1; orange: spectral tuning heater of MCR2; $TE_{0_b}$: fundamental transverse-electric mode in bus waveguide; $TE_{0_r}$: fundamental transverse-electric mode in resonator; $TE_1$: 1st-order transverse-electric mode in resonator). **b** Operation schematic of the bandgap

engineering approach for independent wavelength ($\lambda_r$) and intensity ($I_r$) tuning. The tuning operation is realized by applying voltages to electrical ports $P_1$ and $P_2$, which correspond to microheaters for global and localized thermal tuning of MCRs, respectively. **c** Simulated transmission spectra of the proposed DC-FPR. **d** Electric field distribution in DC-FPR at resonant (left) and non-resonant (right) wavelengths. Measured transmission spectra at C-band (**e**), S-band (**f**) and O-band (**g**). Inset: scanning electron microscopy (SEM) image of the AMDC region. **h** Optical microscope image of the fabricated device. **i** Measured spectra showing the wavelength tuning (left) and intensity tuning (right) of the fabricated device in response to voltages applied to electrical ports $P_1$ and $P_2$, respectively.

The proposed DC-FPR uniquely provides independent, continuous tuning of both wavelength and intensity while maintaining FSR-free operation and ultra-low loss. Wavelength tuning is achieved by applying a voltage to port $P_1$, which induces global heating and causes a synchronous shift in the center wavelengths of both the left and right Bragg gratings (Top panel of Fig. 2b). This, in turn, shifts the resonant wavelength ($\lambda_r$) of the resonator. Intensity modulation is realized by applying power to $P_2$, where localized heating red-shifts only the left grating's center wavelength, reducing the effective overlapping bandwidth ($BW_{overlap}$) and decreasing the reflectivity (bottom panel of Fig. 2b). As a result, the resonant mode intensity ($I_r$) is progressively attenuated, enabling complete extinction. This all-thermal tuning approach offers simpler fabrication and lower optical loss compared to carrier-injection schemes[40-42]. Optical simulations (Fig. 2c) demonstrate single-mode operation over a bandwidth exceeding 300 nm, with low-loss resonant peaks exhibiting an extinction ratio of ＞30 dB and add-port crosstalk below -19 dB. Sidelobes at shorter wavelengths originate from the $TE_{2_r}$ mode but are suppressed to -32. 6 dB due to operation away from critical coupling. As illustrated in Fig. 2d, at the resonant wavelength, energy is confined within the resonator, and simulations confirm low through-port loss at 1.3 μm and 1.6 μm. The longer wavelength exhibits slightly higher coupling losses, which can be mitigated in a cascaded array by placing longer-wavelength channels upstream and using larger coupling gaps. Fabricated devices operating in the O-C bands each exhibit only a single resonant peak within the measurable range, with insertion loss below 0.5 dB (Figs. 2e-2g). By adjusting the coupling gap and the MCR tooth width, the 3-dB bandwidth can be flexibly controlled. We have obtained a 3-dB bandwidth spanning 15-118 GHz, with an insertion loss of less than 1 dB, making it suitable for various application scenarios (Supplement Note 3). An integrated microheater enables dynamic modulation, achieving 16.7 nm (tuning efficiency = 0.07 nm/mW) of continuous wavelength tuning and intensity modulation of a single resonance with over 20 dB drop-port power reduction, as shown in Figs. 2h and 2i. Furthermore, Supplement Notes 5 and 6 demonstrate a nearly eightfold improvement in thermo-optic tuning efficiency, enabled by the undercut process, and wafer-scale device performance on a 12-inch standard 220-nm silicon-on-insulator (SOI) multi-project wafer (MPW).

Capitalizing on these advances in FSR-free bandwidth (>300 nm), low insertion loss (<0.5 dB), and efficient spectral tunability of DC-FPRs, a deeply cascaded architecture can be constructed to address the challenge of achieving high scalability and reconfigurability for versatile on-chip WDM functions.

**High-capacity on-chip WDM communication fabric for optical interconnects and PON**

To directly demonstrate the platform's capability as a high-capacity on-chip WDM communication fabric, we first configured the cascaded DC-FPR network as a single-link wavelength (de)multiplexer ((DE)MUX). In WDM systems, the achievable transmission capacity is fundamentally governed by the performance of (DE)MUX. Leveraging the high scalability of the proposed DC-FPR unit, we constructed a 20-channel DEMUX as plotted in Fig. 3a. To suppress inter-channel crosstalk and allow headroom for higher single-channel data rates, the channel spacing is set to 400 GHz, resulting in a total spectral coverage of 61.3 nm. The fabricated array exhibits insertion losses below 2.5 dB and channel crosstalk below -25 dB across all ports (Fig. 3b). Owing to its FSR-free nature and excellent structural scalability, further increasing the number of channels is straightforward.

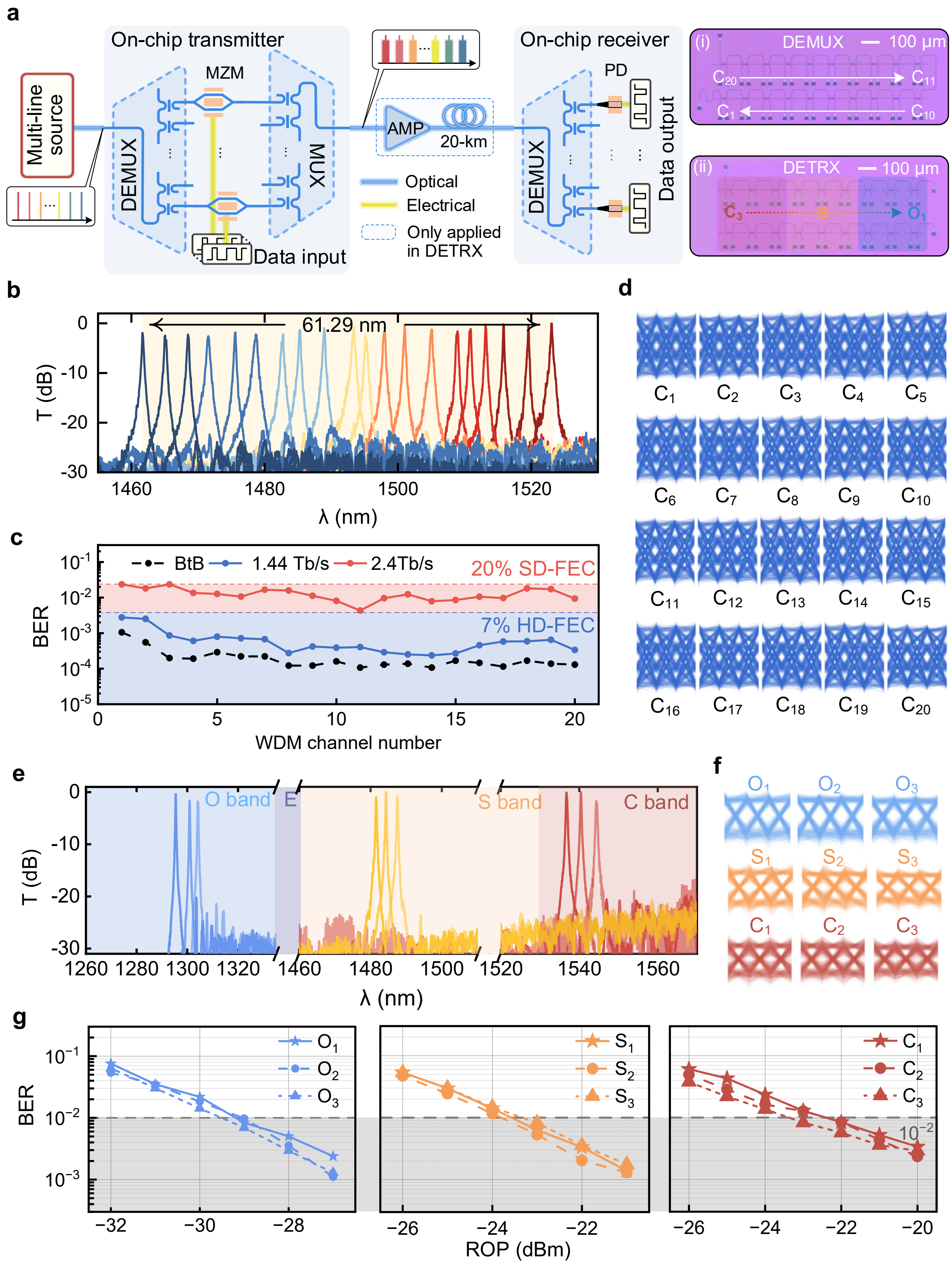

**Fig. 3 Massively multiplexed on-chip WDM communication systems. a** Schematic of the highly-parallelled on-chip WDM communication scheme, along with micrographs of the fabricated DEMUX (i)

and DETRX (ii). (MZM: Mach-Zehnder Modulator; AMP: optical amplifier; PD: photodetector;) **b** Transmission spectra of the 20-channel DEMUX at drop ports from $C_1$ to $C_{20}$. **c** BER performance for each channel under 36-Gbaud and 60-Gbaud PAM4 signal transmission, with back-to-back (BtB) reference without DEMUX shown for 36 Gbaud. Total aggregate data rates reach 1.44 Tbps and 2.4 Tbps, respectively. **d** Eye diagrams of recovered 36-Gbaud PAM4 signals in each channel. **e**. Transmission spectra of the DETRX across the O-to-C band. **f** Eye diagrams of retrieved 100-Gbps (3 × 33.5-Gbaud) NRZ signals in each sub-band channel. **g** BER versus ROP for 33.5-Gbaud NRZ signals in each channel, evaluated near the $1\times10^{-2}$ BER threshold.

To evaluate the system's high-throughput performance, we constructed an experimental system as presented in Fig. 3a. The specific setup description is provided in the Methods section, with further details on the configuration available in Supplement Note 7. Using the measured 3-dB (approximately 36 GHz) and 6-dB (approximately 60 GHz) bandwidths of the DEMUX channel, we have successfully demonstrated parallel transmission of 36-Gbaud PAM4 signals. As illustrated in Fig. 3c, the transmission system achieves a -10-dBm received optical power (ROP) sensitivity per channel, considering a 7% hard-decision forward error correction (HD-FEC) threshold. The system achieves total data rates of 1.44 Tbps and 2.4 Tbps in an intensity modulation and direct detection link. Higher data rates are, in principle, achievable if employing modulators with higher EO bandwidth in this experiment. The performance degradation observed at shorter wavelengths is primarily attributed to the deviation from the central wavelength of the grating coupler and the semiconductor optical amplifier's gain spectrum, as verified in subsequent back-to-back (BtB) tests without DEMUX. Furthermore, each 60-Gbaud PAM4 signal transmission channel maintains a bit error rate (BER) below the 20% soft-decision FEC (SD-FEC) threshold at a ROP of -2 dBm, demonstrating the architecture's potential for high-capacity transmission. The eye diagrams of recovered 36-Gbaud PAM4 signals transmitted over all 20 channels show clear eye opening, as depicted in Fig. 3d.

Beyond single-band multiplexing, we reconfigure the filter responses to realize a multi-subband (de)triplexer (DETRX) capable of concurrent O- and C-band operation, thereby demonstrating its applicability to multi-service scenarios such as next-generation PONs. Expanding on this, we integrate WDM within each service band to create a 9-channel ultra-broadband WDM DETRX with a total O-to-C band coverage exceeding 300 nm. This array achieves insertion loss below 2 dB and an extinction ratio beyond 20 dB across all channels. Transmission tests are conducted with 100-Gbps NRZ signals across all Gigabit-PON bands. The observed ROP variation between bands primarily originated from differences in chromatic dispersion along the 20-km standard single-mode fiber (SSMF). This demonstration of on-chip seamless signal aggregation over more than 300 nm represents a paradigm shift in optical access networks. With the technical advantage of multi-band integration within a single module, this architecture provides a foundation for an "any wavelength to the home" network capable of dynamic bandwidth allocation, aiming to significantly enhance network efficiency and support future broadband services.

**Multi-functional programmable optical processor for dynamic wavelength management**

Beyond multiplexing capabilities, contemporary on-chip WDM systems require essential spectral reconstruction functions, including encoding, splicing, and routing. However, existing integrated optical processors are constrained by limited operational bandwidth and tuning flexibility. To address this, we design and fabricate a 9-channel programmable optical processor suitable for multiple application scenarios, as presented in Fig. 4a (the schematic illustrates a wavelength-selective switch as an application example). By independently programming the resonant and amplitude states of these cascaded units, we dynamically configure the platform to operate as a fully reconfigurable spectral encoder, wavelength scanner, and hitless wavelength-selective switch (WSS), separately. The through port of each preceding channel is linked to the input port of the next, and its drop port is connected to the add port of the subsequent channel. Each channel allows independent control of wavelength and intensity. An optical microscope image of the packaged processor with electrical connections is shown in Fig. 4b. By independently adjusting the intensity of each channel, dynamic spectral encoding can be achieved, enabling the generation of predefined patterns. As illustrated in Fig. 4c, histogram-based triangular and "W"-shaped profiles successfully demonstrated 9-bit binary encoding. Furthermore, when operated as a broadband wavelength scanner, the proposed architecture preserves substantial bandwidth redundancy between adjacent channel wavelengths to ensure complete coverage of the operating bandwidth, ultimately achieving a spectral coverage exceeding 64 nm (Fig. 4d). Given that each device provides a tuning capability of more than 16.7 nm, careful configuration enables an overall scanning range exceeding 150 nm, enabling hitless routing of nine arbitrary-wavelength channels across this continuous band.

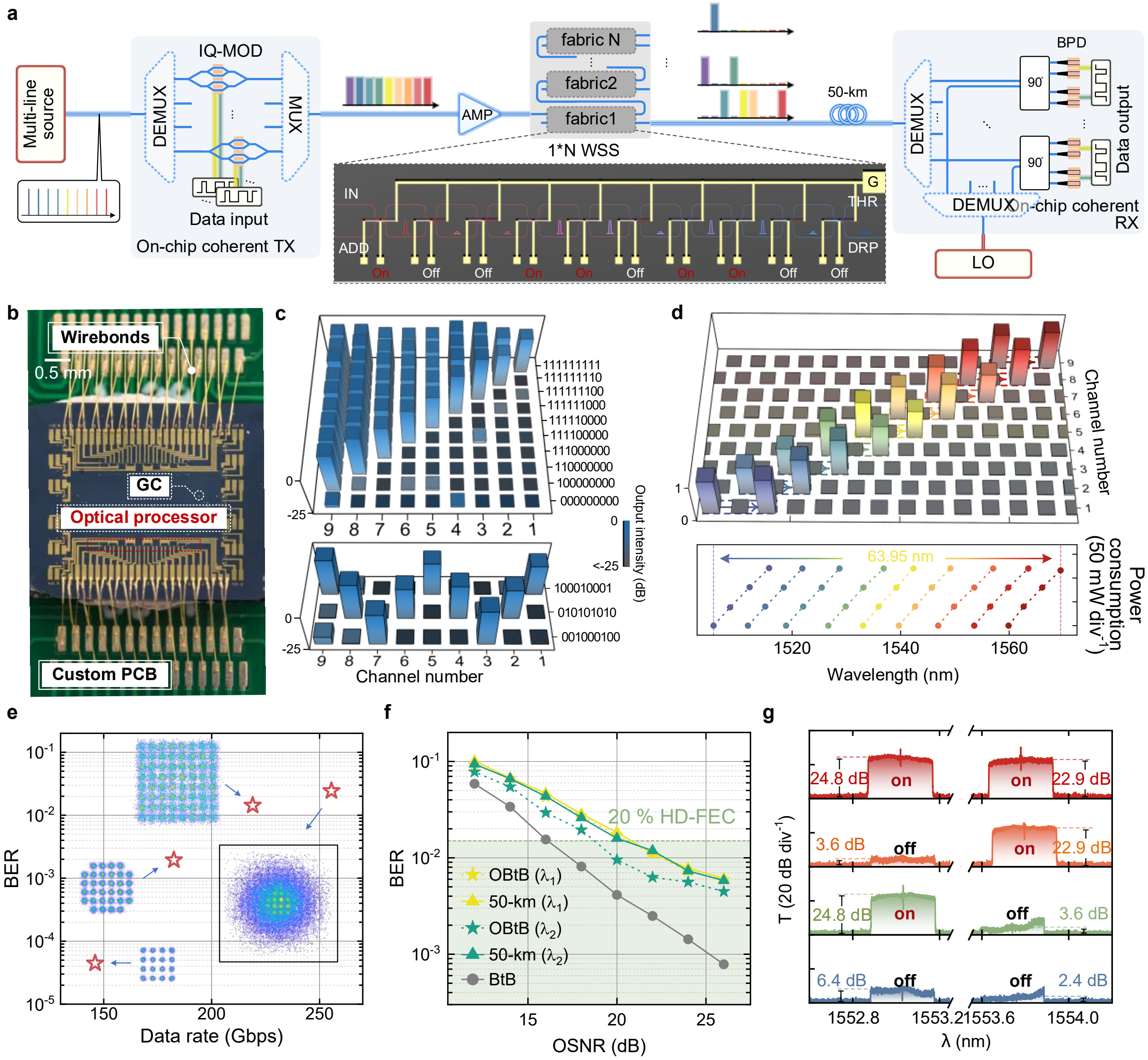

**Fig. 4 9-channel multi-functional optical processor. a** Schematic of the processor architecture and its WSS scheme in a WDM coherent communication system. The fabricated processor consists of nine cascaded units, with the add port of each connected to the drop port of the next. By independently configuring the wavelength and intensity configuration of each channel, the input spectrum can be sliced, modulated, and recombined at the drop port. In this WDM coherent optical communication link, the processor performs channel routing on coherently modulated multi-wavelength spectra, thereby working

as a WSS. (TX: transmitter; RX: receiver; BPD: balanced photodetector; IQ-MOD: in-phase and quadrature-modulator; LO: local oscillator). **b** Photograph of the packaged integrated optical processor. **c** Programmable intensity distribution of each channel acting as a spectral encoder. Top: triangular profile. Bottom: "W"-shaped profile. **d** Normalized transmission response of the 9-channel processor functioning as a wavelength scanner with relayed tuning. By employing spectral stitching, the tuning range of a single channel is extended to an aggregated multi-channel bandwidth. **e** BER versus data rate with recovered constellations under different modulation formats. **f** BER results of dual-wavelength 36.5-Gbaud 32QAM signal co-transmission in optical back-to-back (OBtB), 50-km SSMF transmission through the WSS and BtB link without the WSS under varied OSNR. **g** Modulated spectra of 32QAM signals on the carriers under different channel switching states.

The absence of a periodic FSR enabled truly flexible channel definition and non-blocking operation, validating the platform's role as the core for broadband, reconfigurable optical add-drop multiplexers (ROADMs) and for delivering high-spectral-efficiency WSS. Using adjacent channels, we demonstrate dual-wavelength coherent transmission in a high-throughput configuration (see Supplement Note 8 for the experimental setup). To evaluate spectral efficiency, the transmission quality of quadrature amplitude modulation (QAM) signals has been tested at a symbol rate of 36.5 Gbaud, and the BERs of 16/32/64-QAM signals all remained below the 20% HD-FEC (Fig. 4e). With probabilistic constellation-shaped (PCS) 256-QAM (entropy = 7 bits/symbol), the data rate per channel can achieve 256 Gbps. By reducing the channel spacing to 100 GHz, spectral utilization within a limited bandwidth is effectively improved. To analyze inter-channel crosstalk, closely spaced dual-wavelength signals have been characterized in both back-to-back and 50-km SSMF transmission scenarios. As presented in Fig. 4f, 32-QAM signal transmissions achieve 21-dB of optical signal-to-noise ratio (OSNR) sensitivity with 20% HD-FEC threshold. Leveraging the device's hitless tuning capability, Fig. 4g demonstrates an OSNR contrast exceeding 20 dB between the on and off states of a channel, resulting in an orders-of-magnitude improvement in BER and highlighting excellent switching performance.

**Parallel photonic accelerator for high-throughput matrix operations**

The architecture's independent control over multiple wavelength channels makes it an ideal accelerator for parallel photonic computing, enabling the construction of optical convolution cores that fully exploit the efficiency gains of a WDM platform. To benchmark the performance of our scalable DC-FPR building block, we demonstrate two core convolution tasks: optical edge detection and image classification. Figure 5a shows the fabricated convolution core, featuring a 16-channel cascaded configuration in which the add port of one unit is connected to the drop port of the next. Under static conditions, the resonant wavelengths of the even-numbered channels align with those of the preceding odd-numbered channels. Through thermal tuning via heaters $H_1$-$H_{16}$, the resonant wavelengths of all channels can be precisely aligned to a multi-wavelength laser source with a channel spacing of 1.5 nm. The initially measured transmission spectra are presented in Fig. 5b. Owing to variations in coupling-grating bandwidths and fabrication inconsistencies, the spectral intensities across channels were non-uniform. To mitigate this, a spectral equalization technique was applied to flatten the response, yielding the calibrated spectrum. Each weight unit supports 32 distinct optical intensity levels, corresponding to 5-bit precision, achieved via multi-level modulation, as illustrated in Fig.5c. This precision meets typical quantization accuracy requirements for convolutional neural networks.

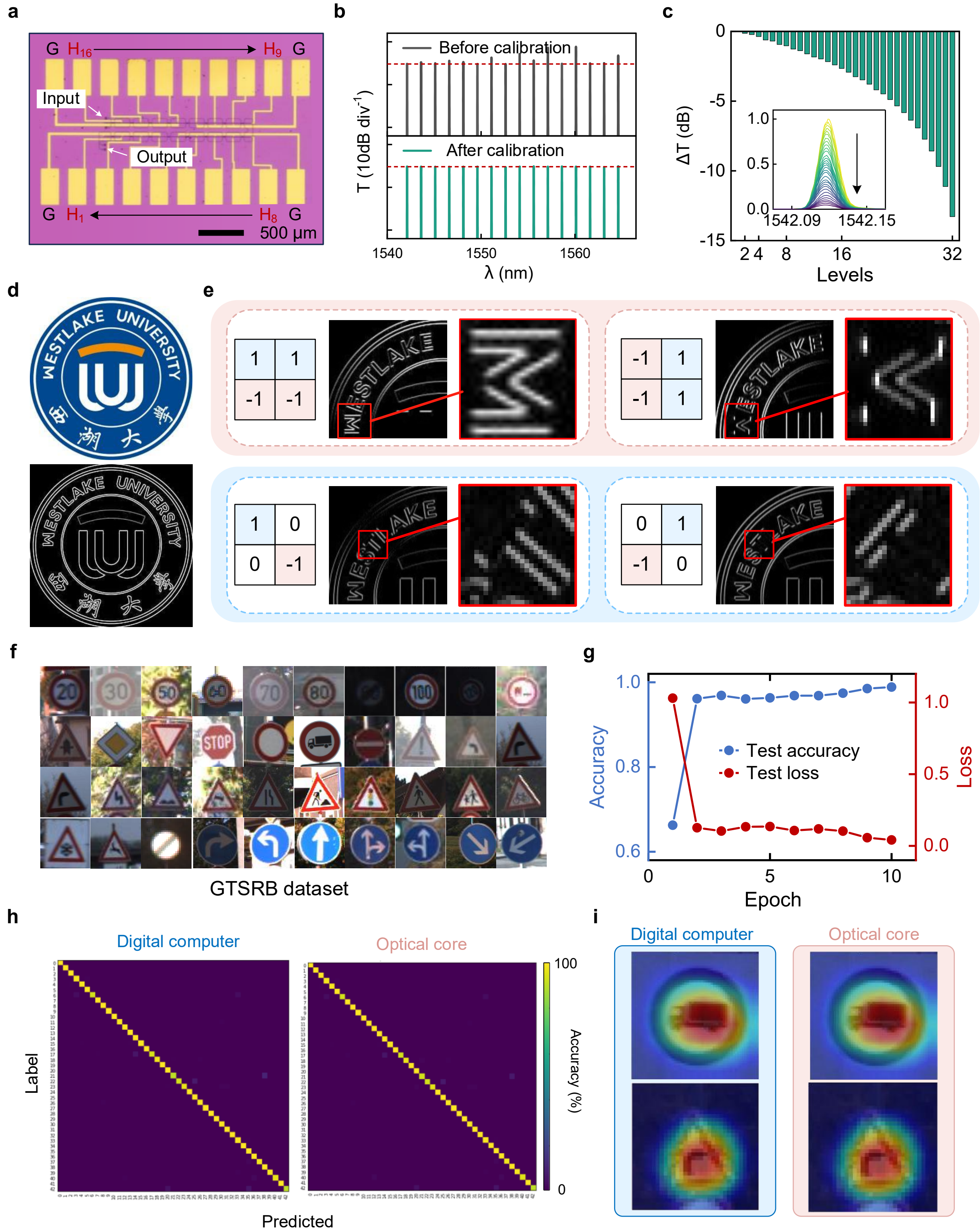

**Fig. 5 16-channel photonic accelerator. a** Optical microscope image of the fabricated 16-channel photonic accelerator. **b** Aligned (top) and calibrated (down) output spectra under multi-Wavelength laser. **c** Distinct calibrated weight levels implemented by the accelerator. The inset shows corresponding spectral responses for each level. **d** Processed images of the Westlake University logo: color version (top) and edge-extracted

result (bottom). **e** Convolutional kernels and corresponding processed outputs for edge detection along horizontal, vertical, 45 degrees, and 135 degrees edge detection. **f** Example images of the GTSRB traffic sign dataset. **g** Classification accuracy and loss results using optical convolution core. **h** Confusion matrices of the convolutional model without (left) and with (right) the optical convolution core on the test of the GTSRB dataset. (row-normalized). **i** Comparisons of the heatmaps for the "No heavy vehicles" and "Sharp right turn ahead" signs obtained using the digital computer (left panel) and the optical convolution core (right panel).

To validate the edge detection functionality, we process an image of the Westlake University logo using the optical core. The logo is converted into an 8-bit grayscale pixel matrix, flattened into a one-dimensional vector, and convolved with four 2×2 kernels based on the Robert operator. These kernels are implemented to extract horizontal, vertical, 45°, and 135° directional edges, respectively. The resulting edge maps are displayed in Fig. 5e, where edges in all orientations are clearly discernible. Moreover, applying a root sum of squares operation to the horizontal and vertical edge outputs enables reconstruction of the complete edge profile, effectively highlighting the edge structures of the image, as shown at the bottom of Fig. 5d. Furthermore, to validate the core's capability in a more complex scenario, we conduct an image classification task on the German Traffic Sign Recognition Benchmark (GTSRB) dataset. The GTSRB dataset comprises 51839 64×64-pixel RGB images categorized into 43 classes, with a split of 35,288 for training, 3,921 for validation, and 1,2630 for testing. Using a PyTorch-based Residual Network as the baseline and comparing it with the results of the optical core. The trained weights from this digital model were mapped onto the optical core. After 10 epochs of training, the classification accuracy and loss evolution of the optical convolution network are depicted in Fig. 5g. Evaluation on the test dataset shows that the network achieves a recognition accuracy of 98.82 %, closely approaching the software baseline accuracy of 98.88%. The row-normalized confusion matrices for both the digital and optical convolution kernels (Fig. 5h) reveal consistently high accuracy for all 43 classes. As plotted in Fig. 5i, Gradient-weighted Class Activation Mapping (Grad-CAM) is applied to the feature maps from the last residual block to generate heatmaps. The feature heatmaps for the "No heavy vehicles" and "Sharp right turn ahead" categories illustrate that the optical kernel effectively captures the features of the traffic signs. With a sustained symbol rate of 40 Gbaud per channel, compatible with standard SOI MPW modulators[43] , the architecture supports a theoretical computational throughput of 1.28 TOPS. Further details are available in the Supplement Note 9. This result establishes a clear, scalable path to leveraging the platform as a programmable optical accelerator.

## Discussion

This paper presents the experimental validation of an add-drop bandpass filter utilizing a DC-FPR architecture. The device incorporates asymmetric multimode Bragg gratings to enable efficient mode conversion, resulting in high energy-extraction efficiency at the drop port of the standing-wave resonator. Through band engineering techniques, the device achieves calibration-free and FSR-free operation. This approach also provides flexible wavelength and intensity tuning with low optical loss, without requiring complex fabrication steps. Table 1 summarizes representative works on resonators with large FSR to date. The proposed DC-FPR structure achieves exceptional performance, featuring low loss (<0.5 dB), FSR-free operation (>300 nm), calibration-free tuning, and flexible control. These properties make it an ideal building block for a programmable, scalable on-chip WDM processing platform for a range of applications.

In a record-setting demonstration, a single-link cascade of 20 wavelength channels, operating without external tuning, achieves a channel spacing of around 400 GHz, with insertion loss as low as 2.5 dB and crosstalk suppressed below -25 dB (Supplementary Table 2). These outcomes underscore the device's strong potential as a foundational component in future high-capacity, large-scale WDM systems.

Further extending its functionality, an optical processor based on this architecture accomplishes 9-bit binary encoding using histogram-based triangular and "W"-shaped spectral profiles. The same structure also operates as a continuous-wavelength scanner. By exploiting silicon's high thermo-optic coefficient, each tuning unit covers 16.7 nm, enabling aggregate spectral coverage beyond 150 nm. In photonic computing applications, the platform supports not only latency-sensitive tasks, such as edge image recognition, but also provides essential hardware underpinnings for general-purpose photonic accelerators and large-scale optical neural networks. Operating at a symbol rate of 40 Gbaud, the system delivers a computational throughput of 1.28 TOPS. In contrast to other multi-dimensional multiplexing approaches, the architecture markedly reduces reliance on intricate synchronization and data-acquisition mechanisms. At the same time, its add-drop-based weight bank natively supports complex-valued computations.

As a versatile platform technology, the device offers broad applicability across optical communications, signal processing, and photonic computing, with the potential to advance WDM systems toward greater capacity and scalability. Employing established wafer-level silicon photonic manufacturing, along with ultra-broadband low-loss fiber-to-chip couplers and tighter channel spacing, single-link operation supporting beyond 80 WDM channels appears feasible (insertion loss ~10 dB) (Supplement Note 4). In combination with hybrid wavelength-, mode-, and polarization-division multiplexing over a 7-core fiber using advanced modulation at 200 Gbps per channel, Pbps-scale link transmission capacities become attainable, thereby enabling the realization of 256 TOPS photonic convolution core (Supplement Note 4).

**Table 1| Performance comparisons of on-chip large-FSR/FSR-free devices**

| Structure | FSR (nm) | Filter type | $IL_{drop}$ (dB) | Wav. /Int. | FSR Cali. (Y/N) | $ER_{drop}$ (dB) | Size (μm × μm) | Refs |
|---|---|---|---|---|---|---|---|---|
| High-order MRR | 37 | Add-drop | 0.2 | Wav. | N | 35 | 8×16 | [44] |
| MZI-assisted 4-order MRR | 90 | Add-drop | 1 | Both | Y | 35 | 190×50 | [37] |
| CDC-assisted MRR | 60 | Add-drop | 1 | Wav. | Y | 18 | 1963 | [36] |
| Low-loss AWG | 29 | Fan-in/out | 2.2 | N.A. | N | ~30 | 600×800 | [31] |
| MZI-assisted AWG | 28.8 | Fan-in/out | 8 | N.A. | N | 20-30 | 520×190 | [33] |
| MBG | >100 | Drop | 1 | N.A. | N | 27 | 500×40 | [45] |
| Nanobeam FPR | >300 | All-pass | N.A. | Wav. | N | N.A. | 10×5 | [46] |
| SM-FPR | >300 | Add-drop | 6.5 | Wav. | N | 21.5 | 150×25 | [39] |
| MM-FPR | 40 | Add-drop | 0.6 | Wav. | N | ~25 | 500×5 | [47] |
| MM-FPR | >100 | Add-drop | 0.7 | Wav. | N | 19.8 | 300×50 | [48] |
| **DC-FPR** | **>300** | **Add-drop** | **<0.5** | **Both** | **N** | **~25** | **150×10** | **This work** |

In column "FSR Cali.", the requirement of active control components for large FSR/FSR-free spectrum calibration per resonator is indicated. Acronyms: (MZI) Mach-Zehnder interferometer, (CDC) contra-directional coupler, (AWG) arrayed-waveguide grating, (MBG) multimode Bragg grating (SM) single mode, (MM) multimode, (FPR) F-P resonator, ($IL_{drop}$) insertion loss at drop port, (Wav. /Int.) wavelength/intensity, (N.A.) Not available, (Y/N) yes/no, ($ER_{drop}$) extinction ratio of drop port.

# References

1 Bunandar, D. Passage M1000 : A 3D Photonic Interposer for AI. *2025 IEEE Hot Chips 37 Symposium (HCS)*. (2025).

2 Hua, S. *et al.* An integrated large-scale photonic accelerator with ultralow latency. *Nature* **640**, 361-367, doi:10.1038/s41586-025-08786-6 (2025).

3 Daudlin, S. *et al.* Three-dimensional photonic integration for ultra-low-energy, high-bandwidth interchip data links. *Nature Photonics* **19**, 502-509, doi:10.1038/s41566-025-01633-0 (2025).

4 Gholami, A. *et al.* AI and Memory Wall. *IEEE Micro* **44**, 33-39, doi:10.1109/mm.2024.3373763 (2024).

5 Jun Shan Wey. The Outlook for PON Standardization: A Tutorial. *Journal of Lightwave Technology* **38**, 31-42, doi:10.1109/jlt.2019.2950889 (2020).

6 Abbas, H. S. & Gregory, M. A. The next generation of passive optical networks: A review. *Journal of Network and Computer Applications* **67**, 53-74, doi:10.1016/j.jnca.2016.02.015 (2016).

7 Liu, Y. *et al.* Parallel wavelength-division-multiplexed signal transmission and dispersion compensation enabled by soliton microcombs and microrings. *Nat Commun* **15**, 3645, doi:10.1038/s41467-024-47904-2 (2024).

8 Zhang, C. *et al.* Silicon Photonic Wavelength-Selective Switch Based on an Array of Adiabatic Elliptical-Microrings. *J. Lightwave Technol.*, 1-8, doi:10.1109/jlt.2023.3264613 (2023).

9 Wang, Y. *et al.* Reconfigurable versatile integrated photonic computing chip. *eLight* **5**, 20, doi:10.1186/s43593-025-00098-6 (2025).

10 Chang, L., Liu, S. & Bowers, J. E. Integrated optical frequency comb technologies. *Nature Photonics* **16**, 95-108, doi:10.1038/s41566-021-00945-1 (2022).

11 Kippenberg, T. J., Holzwarth, R. & Diddams, S. A. Microresonator-Based Optical Frequency Combs. *Science* **332**, 555-559 (2011).

12 Trocha, P. *et al.* Ultrafast optical ranging using microresonator soliton frequency combs. *Science* **359**, 887-891 (2018).

13 A. A. Jørgensen, D. Kong, M. R. Henriksen & F. Klejs. Petabit-per-second data transmission using a chip-scale microcomb ring resonator source. *Nature Photonics* **16**, 798-802, doi:10.1038/s41566-022-01082-z (2022).

14 Marin-Palomo, P. *et al.* Microresonator-based solitons for massively parallel coherent optical communications. *Nature* **546**, 274-279, doi:10.1038/nature22387 (2017).

15 Yang, K. Y. *et al.* Multi-dimensional data transmission using inverse-designed silicon photonics and microcombs. *Nat Commun* **13**, 7862, doi:10.1038/s41467-022-35446-4 (2022).

16 Corcoran, B. *et al.* Optical microcombs for ultrahigh-bandwidth communications. *Nature Photonics* **19**, 451-462, doi:10.1038/s41566-025-01662-9 (2025).

17 Bai, B. *et al.* Microcomb-based integrated photonic processing unit. *Nat Commun* **14**, 66, doi:10.1038/s41467-022-35506-9 (2023).

18 Xu, X. *et al.* 11 TOPS photonic convolutional accelerator for optical neural networks. *Nature* **589**, 44-51, doi:10.1038/s41586-020-03063-0 (2021).

19 Shu, H. *et al.* Microcomb-driven silicon photonic systems. *Nature* **605**, 457-463, doi:10.1038/s41586-022-04579-3 (2022).

20 Zhao, W. *et al.* 96-Channel on-chip reconfigurable optical add-drop multiplexer for multidimensional multiplexing systems. *Nanophotonics* **11**, 4299-4313, doi:10.1515/nanoph-2022-0319 (2022).

21 Peng, Y., Zhao, W., Shi, Y. & Dai, D. 192-channel silicon Reconfigurable Optical Add-Drop Multiplexer. *IEEE Silicon Photonics Conference.* **TuP20** (2024).

22 Yuan, Y. *et al.* A 5 x 200 Gbps microring modulator silicon chip empowered by two-segment Z-shape junctions. *Nat Commun* **15**, 918, doi:10.1038/s41467-024-45301-3 (2024).

23 Sun, J. *et al.* A 128 Gb/s PAM4 Silicon Microring Modulator With Integrated Thermo-Optic Resonance Tuning. *Journal of Lightwave Technology* **37**, 110-115, doi:10.1109/jlt.2018.2878327 (2019).

24 Eppenberger, M. *et al.* Resonant plasmonic micro-racetrack modulators with high bandwidth and high temperature tolerance. *Nature Photonics* **17**, 360-367, doi:10.1038/s41566-023-01161-9 (2023).

25 Zhou, H. *et al.* Photonic matrix multiplication lights up photonic accelerator and beyond. *Light Sci. Appl.* **11**, 30, doi:10.1038/s41377-022-00717-8 (2022).

26 Pintus, P. *et al.* Integrated non-reciprocal magneto-optics with ultra-high endurance for photonic in-memory computing. *Nature Photonics* **19**, 54-62, doi:10.1038/s41566-024-01549-1 (2024).

27 Liu, S. *et al.* Calibration-free and precise programming of large-scale ring resonator circuits. *Optica* **12**, 1113-1121, doi:10.1364/optica.557415 (2025).

28 Bogaerts, W. *et al.* Silicon microring resonators. *Laser Photon Rev* **6**, 47-73, doi:10.1002/lpor.201100017 (2011).

29 Dae-Won Rho *et al.* Energy Efficient Monolithically Integrated 256 Gb/s Optical Transmitter With Autonomous Wavelength Stabilization in 45 nm CMOS SOI. *IEEE Journal of Solid-State Circuits* **60**, 2522-2531, doi:10.1109/jssc.2024.3511673 (2025).

30 Pirmoradi, A. *et al.* Integrated multi-port multi-wavelength coherent optical source for beyond Tb/s optical links. *Nat Commun* **16**, 6387, doi:10.1038/s41467-025-61288-x (2025).

31 Shen, X. *et al.* Ultra-Low-Crosstalk Silicon Arrayed-Waveguide Grating (De)multiplexer with 1.6-nm Channel Spacing. *Laser Photon Rev* **18**, 2300617, doi:10.1002/lpor.202300617 (2023).

32 Zhong, Y. *et al.* Flat-Top Eight-Channel Wavelength Division Demultiplexer for L-Band Optical Communication With Low Crosstalk. *IEEE Photonics Journal* **17**, 1-6, doi:10.1109/jphot.2025.3607596 (2025).

33 Chen, S. *et al.* Compact Dense Wavelength-Division (De)multiplexer Utilizing a Bidirectional Arrayed-Waveguide Grating Integrated With a Mach-Zehnder Interferometer. *J. Lightwave Technol.* **33**, 2279-2285, doi:10.1109/jlt.2015.2405510 (2015).

34 Feldmann, J. *et al.* Parallel convolutional processing using an integrated photonic tensor core. *Nature* **589**, 52-58, doi:10.1038/s41586-020-03070-1 (2021).

35 Vlasov, Y., Green, W. M. J. & Xia, F. High-throughput silicon nanophotonic wavelength-insensitive switch for on-chip optical networks. *Nature Photonics* **2**, 242-246, doi:10.1038/nphoton.2008.31 (2008).

36 Eid, N. *et al.* FSR-free silicon-on-insulator microring resonator based filter with bent contra-directional couplers. *Opt Express* **24**, 29009-29021, doi:10.1364/OE.24.029009 (2016).

37 Morichetti, F. *et al.* Polarization-transparent silicon photonic add-drop multiplexer with wideband hitless tuneability. *Nat. Commun.* **12**, 4324, doi:10.1038/s41467-021-24640-5 (2021).

38 Sun, C. *et al.* Free-spectral-range-free filters with ultrawide tunability across the S + C + L band. *Photonics Res* **9**, 1013-1018, doi:10.1364/prj.420005 (2021).

39 Sun, C. *et al.* Tunable narrow-band single-channel add-drop integrated optical filter with ultrawide FSR. *PhotoniX* **3**, 12, doi:10.1186/s43074-022-00056-2 (2022).

40 Nishi, H. *et al.* Monolithic Integration of a Silica-Based Arrayed Waveguide Grating Filter and Silicon Variable Optical Attenuators Based on p–i–n Carrier-Injection Structure. *Applied Physics Express* **3**, doi:10.1143/apex.3.102203 (2010).

41 Wang, X. *et al.* Compact silicon photonic resonance-sssisted variable optical attenuator. *Opt Express* **24**, 27600-27613, doi:10.1364/OE.24.027600 (2016).

42 Yuan, P. *et al.* Design and fabrication of two kind of SOI-based EA-type VOAs. *Optics & Laser Technology* **102**, 166-173, doi:10.1016/j.optlastec.2017.12.041 (2018).

43 Ding, R. *et al.* A Compact Low-Power 320-Gb/s WDM Transmitter Based on Silicon Microrings. *IEEE Photonics Journal* **6**, 1-8, doi:10.1109/jphot.2014.2326656 (2014).

44 Liu, D., Zhang, L., Tan, Y. & Dai, D. High-Order Adiabatic Elliptical-Microring Filter with an Ultra-Large Free-Spectral-Range. *J. Lightwave Technol.* **39**, 5910-5916, doi:10.1109/jlt.2021.3091724 (2021).
45 Liu, D., Zhang, M. & Dai, D. Low-loss and low-crosstalk silicon triplexer based on cascaded multimode waveguide gratings. *Opt Lett* **44**, 1304-1307, doi:10.1364/ol.44.001304 (2019).
46 Tang, R. *et al.* High-resolution 2D quasi-distributed optical sensing with on-chip multiplexed FSR-free nanobeam cavity array. *Laser Photon Rev* (2023).
47 Qiu, H. *et al.* Silicon Add-Drop Filter Based on Multimode Bragg Sidewall Gratings and Adiabatic Couplers. *J. Lightwave Technol.* **35**, 1705-1709, doi:10.1109/jlt.2017.2667711 (2017).
48 Chen, Y. *et al.* Narrow-band and FSR-free Add-drop Filter on SOI Based on Side-coupled π-phase-shifted Antisymmetric Bragg Grating Resonator. *J. Lightwave Technol.* **43**, 9258-9269, doi:10.1109/jlt.2025.3594428 (2025).

# Methods

### Device Fabrication and Packaging

The devices are fabricated on 220-nm silicon-on-insulator (SOI) wafers. The fabrication process begins with cleaning the substrate, followed by spin-coating with an electron-beam resist. High-resolution patterning is achieved using 100 kV electron-beam lithography (EBL) and subsequent development. The pattern is then transferred into the silicon device layer by inductively coupled plasma reactive ion etching (ICP-RIE) (Oxford PlasmaPro 100 Cobra), with an etch depth of 150 nm. An 800-nm-thick silicon oxide layer is subsequently deposited via plasma-enhanced chemical vapor deposition (PECVD) (Samco PD-220NL) to serve as a cladding layer, minimizing optical loss from subsequently deposited metal electrodes. Micro-heaters (50 nm Pt/150 nm Au) are then patterned by electron-beam evaporation (ULVAC ei-5z) followed by lift-off. To reduce power consumption in the energy-efficient devices presented in this work, deep trench etching by ICP-RIE, followed by selective undercutting using $XeF_2$ vapor, is performed to create suspended structures for improved thermal isolation. Finally, multi-channel electrical connections between the chip and a custom-designed printed circuit board (PCB) are established by wire bonding.

### Experimental setup for WDM signal quality characterization

In the DEMUX and DETRX characterization, a modulated signal is generated using an arbitrary waveform generator to drive a commercial high-speed electro-optic (E-O) modulator, which encodes the signal onto a designated optical carrier. The modulated light is coupled into the chip via an optical fiber aligned on a high-precision translation stage. Each output channel is detected by a high-speed photodetector (O-E conversion) and recorded by a real-time high-speed oscilloscope. Offline digital signal processing (DSP) is subsequently performed, including waveform generation, match filtering, and post-processing at both transmitter and receiver ends. Optical amplifiers and attenuators are incorporated into the link to evaluate BER performance over a range of received optical powers. For DETRX characterization, optical components are adjusted to the operating wavelength band, and an additional 20-km fiber span was added to the transmission link.

In the WSS characterization, two optical signals are simultaneously modulated on two assigned optical carriers. Similar to the DEMUX tests, the procedure involved offline DSP-based waveform generation, E-O and O-E conversion, and DSP-assisted signal recovery. Optical amplifiers and attenuators were used to compensate for 50-km transmission loss and to measure BER under various conditions. A portion of the output light is also directed to an optical spectrum analyzer (OSA) for OSNR monitoring. Additional optical and electrical amplifiers are employed to compensate for on-chip coupler insertion loss and to enhance electrical signal levels, respectively.

# Data Availability

The data underlying the results presented in this paper are available from the authors upon request.

## Acknowledgements
The authors would like to acknowledge the Westlake Center for Micro/Nano Fabrication and Instrumentation, the Service Center for Physical Sciences at Westlake University, and the ZJU Micro-Nano Fabrication Center at Zhejiang University for their support in providing facilities.

## Funding
"Pioneer" and "Leading Goose" R&D Program of Zhejiang Province (2024SDXHDX0005); National Key Research and Development Program of China (2023YFB2806504); National Natural Science Foundation of China (62175202, 62471404); Key Project of Westlake Institute for Optoelectronics (2023GD003/110500Y0022303); Research Center for Industries of the Future (RCIF) at Westlake University (Grant No. 210230006022302/002).

## Author contribution
B. S. and H. Z. contributed equally to this work. L. L. led the general project study. W. S. led the chip testing for optical communication, while H. L. led the demonstration of the device block for scalable optical convolution processing. B. S. performed the theoretical analysis and device simulation while K. B. contributed to the device model development. B. S., K. L., Y. Y., D. H., Q. D., Z. C., Y. W. and Q. Z. contributed to the chip fabrication and packaging. B. S. and H. Z. contributed the chip tests. Y. S. and W. L. contributed the signal processing code. Y. F. contributed to the characterization of signal quality. Z. W. contributed the convolution computing design. M. W., W. C, and X. S. assisted with wafer-scale device fabrication. B. S., H. Z., Y. S., Z. W., and L. L. prepared the manuscript draft. H. L., W. S., and L. L. supervised and coordinated the project. All authors processed and analyzed the results. All authors contributed to the review and revision of the manuscript.

## Conflict of interest
Authors Maoliang Wei, Wei Cao, and Xu Sun were employed by Crealights Technology Co., Ltd. The remaining authors declare that the research was conducted in the absence of any commercial or financial relationships that could be construed as a potential conflict of interest.

## Supporting Information
Supporting Information is available from the Nature Online Library or the author upon request.